\documentstyle[12pt]{article}
\newlength{\extraspace}
\newlength{\extraspaces}

\newcommand{\newsection}[1]{
\vspace{15mm}
\pagebreak[3]
\addtocounter{section}{1}
\setcounter{subsection}{0}
\setcounter{footnote}{0}
\setcounter{equation}{0}
\begin{flushleft}
{\large\bf \thesection. #1}
\end{flushleft}
\nopagebreak
\medskip
\nopagebreak}

\begin{document}
\addtolength{\baselineskip}{.7mm}

\thispagestyle{empty}

\begin{flushright}
{\sc PUPT}-1458\\
March 1994
\end{flushright}
\vspace{.3cm}

\begin{center}
{\Large\bf{Hawking Radiation from Black Holes
 \\[2mm] Formed During Quantum Tunneling}}\\[20mm]
{\sc Per Kraus}\\[3mm]
{\it Joseph Henry Laboratories\\[2mm]
 Princeton University\\[2mm]
Princeton, NJ 08544\\[2mm]
 E-mail: perkraus@puhep1.princeton.edu}
\\[30mm]

{\sc Abstract}
\end{center}
We study the behaviour of scalar fields on background geometries which undergo
quantum tunneling. The two examples considered are a moving mirror in flat
space which tunnels through a potential barrier, and a false vacuum bubble
which tunnels to form a black hole. WKB approximations to the Schr\"{o}dinger
and Wheeler-DeWitt equations are made, leading one to solve field equations on
the Euclidean metric solution interpolating between the classically allowed
geometries.  The state of the field after tunneling can then be determined
using the method of non-unitary Bogolubov transformations developed by Rubakov.
It is shown that the effect of the tunneling is to damp any excitations
initially present, and, in the case of the black hole, that the behaviour of
fields on the Euclidean Kruskal manifold ensures that the late time radiation
will be thermal at the Hawking temperature.

\noindent

\vfill

\newpage

\newsection{Introduction}

The radiation of particles from matter evolving along a classical trajectory
has been heavily studied in recent years.  Less well studied is the radiation
accompanying quantum tunneling from one classically allowed trajectory to
another.  The following question is of interest: if a matter system impinges
upon a potential barrier with a radiation field in a certain state, what is
the state of the field given that the matter is subsequently observed to be
on the other side of the barrier?  A method to answer this question in the
context of false vacuum decay in flat space was developed by Rubakov
 \cite{Rub84} and has been generalized to include gravity as well as topology
changing processes \cite{Rub87,Kan89}.  The spectrum of radiation is
 found   by solving an imaginary time Schr\"{o}dinger equation, the occurrence
of which leads to novel features. Instead of solving field equations in real
time, one is naturally led to consider propagation on the Euclidean solution
interpolating between the two classical trajectories. As phase factors in real
time are converted into exponential damping factors in imaginary time, the
resulting particle creation can be distinctly different and is accompanied by
the systematic supression of excitations present before tunneling.

Given this situation, it is natural to ask how the radiation from black holes
might be affected by the presence of tunneling.  If we consider a distribution
of matter, initally outside its Schwarzschild radius, which tunnels through a
potential barrier to form a black hole, the conventional calculation
 \cite{Haw75} of the radiation does not apply. On the other hand, it would be
shocking if the same answer was not obtained for the radiation at late times,
as this is thought to depend only on the hole's late time geometry and not on
its history at early times.  Here we compute the radiation for this process
and show that while the Euclidean time evolution has an effect at early times,
it has none at late times so that the standard result is in fact obtained.

In order illustrate the technique of Ref.~\cite{Rub84} in a simpler setting, we
 first study the effect of tunneling on another well known radiating system ---
the moving mirror \cite{Dav77}. We show in Sect.~(2)
 how an imaginary
 time Schr\"{o}dinger equation emerges from a Born-Oppenheimer
approximation, and use this result to calculate the shift in the spectrum of
radiated particles as a result of the tunneling.  It is shown that the
initial spectrum  is shifted to favor low energy excitations,
as is understood by realizing that the probability to tunnel is increased if
energy is transferred from the radiation to the mirror.

In Sect.~(3) this approach is extended to include gravity in
asymptotically flat space.  A WKB approximation to the Wheeler-DeWitt equation,
as considered in \cite{Lap79,Banks85}, is used to obtain an imaginary time
Schr\"{o}dinger equation which can then be solved as before. In
 Sect.~(4) we use this result to examine the radiation from a
black hole which is formed by tunneling.  In particular, we consider the
tunneling of a false vacuum bubble, a system extensively studied in
 Refs.~\cite{Sato81} ---~\cite{Guth87}.  This example involves a
 complication due to the peculiar structure that arises;
 Refs.~\cite{Guth90,Pol90} show that the sequence of
three-geometries encountered during tunneling can not be stacked together to
form a manifold.  Employing a slight modification of the
standard approach, we show how the behaviour of fields on the Euclidean
Schwarzschild manifold protects the late time radiation from being affected
by tunneling.  An intuitive reason for this is that the bubble's tunneling
 probability
is unchanged by the presence of Hawking radiation, which involves the creation
of pairs of particles with zero total energy.

\newsection{Tunneling Mirror}

\label{secmir}
Consider a mirror moving in a one dimensional potential in the presence of a
massless scalar field.  The Schr\"{o}dinger equation for this system is
\begin{equation}
  [\hat{H}_{m}+\hat{H}_{\phi}]\Psi[\phi,x_{m};t]=i\frac{\partial}{\partial
  t}\Psi[\phi,x_{m};t]
\end{equation}
where
\begin{equation}
\hat{H}_{m}=-\frac{1}{2m}\frac{\partial^{2}}{\partial x_{m}^{\;2}}+V(x_{m})
\end{equation}
and
\begin{equation}
\hat{H}_{\phi}=\frac{1}{2}\int_{x_{m}}^{\infty}dx\left[-\frac{\delta^{2}}
{\delta\phi(x)^{2}}+\left(\frac{d\phi}{dx}\right)^{2}\right].
\end{equation}
Note that $\Psi$ is a function of the mirror coordinate $x_{m}$, and a
functional of the field configuration $\phi(x)$.  The mirror boundary
condition is imposed by demanding that the field vanish at $x_{m}$,
\begin{equation}
  \Psi[\phi,x_{m};t]=0\; \mbox{ \ if \ } \phi(x_{m})\neq0.
\end{equation}
The system is solved by assuming that the backreaction of the field on the
mirror is a small perturbation of the mirror's motion, and that the mass and
momenta of the mirror are large enough that it can be described by a well
localized wave packet.  In this domain the system admits a Born-Oppenheimer
approximation, which amounts to an expansion in $1/m$.  In particular, we seek
a solution to the time independent Schr\"{o}dinger equation
\begin{equation}
[\hat{H}_{m}+\hat{H}_{\phi}]\Psi[\phi,x_{m}]=E\Psi[\phi,x_{m}]
\label{e:ti}
\end{equation}
valid to zeroth order in $1/m$.  Following Refs.~\cite{Rub84,Banks85} the
Born-Oppenheimer approximation is implemented by writing $\Psi$ in the form
\begin{equation}
\Psi[\phi,x_{m}]=\psi_{VV}(x_{m})\,e^{iS(x_{m})}\,\chi[\phi,x_{m}]
\end{equation}
where $\psi_{VV}$ is a slowly varying function to be identified with the
Van Vleck determinant.  To lowest order in $1/m$, (\ref{e:ti}) reduces to
the Hamilton-Jacobi equation.
\begin{equation}
\frac{1}{2m}\left(\frac{dS}{dx_{m}}\right)^{2}+V(x_{m})=E
\label{e:hj}
\end{equation}
since $dS/dx_{m}$, $V(x_{m})$ and $E$ are all of order $m$.

To zeroth order:
\begin{equation}
-\frac{i}{2m}\,\frac{d^{2}S}{dx_{m}^{\;2}}\,\psi_{VV}\,\chi[\phi,x_{m}]
-\frac{i}{m}\frac{dS}{dx_{m}}\frac{d\psi_{VV}}{dx_{m}}\chi[\phi,x_{m}]
\end{equation}
$$
-\frac{i}{m}\psi_{VV}\frac{dS}{dx_{m}}\frac{\partial}{\partial
x_{m}}\chi[\phi,x_{m}]
  +  \psi_{VV}\,\hat{H}_{\phi}\,\chi[\phi,x_{m}]=0.
$$
$\psi_{VV}$ is chosen so that the first two terms cancel, leaving
\begin{equation}
\hat{H}_{\phi}\, \chi[\phi,x_{m}]=\frac{i}{m}\frac{dS}{dx_{m}}
\frac{\partial}{\partial x_{m}}\chi[\phi,x_{m}].
\end{equation}
This can be put in a familiar form by defining the time variable $\tau(x_{m})$.
 In a classically allowed region, where $E-V(x_{m})>0$ and $dS/dx_{m}$ is real,
$ \tau $ is defined by
\begin{equation}
\frac{d\tau}{d x_{m}} = \frac{m}{dS/dx_{m}}
\mbox{ \ \ \ allowed regions}
\end{equation}
whereas in a classically forbidden region with  $dS/dx_{m}$ imaginary,
\begin{equation}
\frac{d\tau_{E}}{dx_{m}} = i \frac{m}{dS/dx_{m}}
\mbox{ \ \ \ forbidden regions.}
\end{equation}
The resulting zeroth order equations for $\phi$ are:
\begin{equation}
\hat{H}_{\phi} \,\chi[\phi,\tau] = i \frac{\partial}{\partial \tau}\chi[\phi,
\tau] \mbox{ \ \ \ allowed regions}
\end{equation}
\begin{equation}
-\hat{H}_{\phi} \,\chi[\phi,\tau_{E}] = \frac{\partial}{\partial \tau_{E}}
\chi[\phi,\tau_{E}] \mbox{ \ \ \ forbidden regions.}
\end{equation}
These are the fundamental equations governing the evolution of the scalar
field in the presence of the mirror.  In the allowed regions we have
recovered the time-dependent Schr\"{o}dinger equation with the postion of the
mirror playing the role of a clock, whereas in the forbidden regions we have
obtained a diffusion equation, which we shall refer to as the Euclidean
Schr\"{o}dinger equation, with the Euclidean time $\tau_{E}$ measuring the
position of the mirror in the potential barrier.

Now, choose the potential to be of the form illustrated in Fig. 1 and let the
mirror come from right to left.  In the allowed region to the right of
 $x_{m}^{i}$
 the state $\chi[\phi,\tau]$ obeys the normal Schr\"{o}dinger equation, and so
standard methods can be used to find $\chi[\phi,\tau^{i}]$.  Between
$x_{m}^{i}$ and $x_{m}^{f}$ the mirror is in a forbidden region, so the
state evolves according to
\begin{equation}
-\frac{1}{2}\int_{x_{m}(\tau_{E})}
^{\infty} dx \left[-\frac{\delta^{2}}{\delta \phi(x)^{2}}
+ \left(\frac{d\phi}{dx}\right)^{2}\right] \chi[\phi,\tau_{E}]
=\frac{\partial}{\partial \tau_{E}} \chi[\phi,\tau_{E}]
\end{equation}
with $\chi[\phi,\tau_{E}^{i}]=\chi[\phi,\tau^{i}]$.
We wish to solve this equation  in order to find the state at the final turning
point $x_{m}^{f}$. It is useful to transform the mirror to rest by defining the
coordinate
\begin{equation}
y(x,\tau_{E})=x-x_{m}(\tau_{E})
\end{equation}
in terms of which the Euclidean Schr\"{o}dinger equation is
\begin{equation}
-\frac{1}{2} \int_{0}^{\infty} dy \left[ -\frac{\delta^{2}}{\delta \phi(y)
^{2}} + 2\,\frac{dx_{m}}{d\tau_{E}} \frac{d\phi}{dy}\frac{\delta}{\delta
\phi(y)} + \left(\frac{d\phi}{dy}\right)^{2}\right] \chi[\phi,\tau_{E}]
=\frac{\partial}{\partial \tau_{E}} \chi[\phi,\tau_{E}]
\end{equation}
or
\begin{equation}
-\hat{H}_{\phi}^{E}(\tau_{E}) \chi[\phi,\tau_{E}] =
\frac{\partial}{\partial \tau_{E}} \chi[\phi,\tau_{E}].
\end{equation}
The solution is
\begin{equation}
\chi[\phi,\tau_{E}]=T\exp\left[-\int_{\tau_{E}^{i}}^{\tau_{E}}
\hat{H}_{\phi}^{E}(\tau_{E}^{'}) d\tau_{E}^{'}\right] \chi[\phi,\tau_{E}^{i}]
=\hat{U}_{E}(\tau_{E},\tau_{E}^{i})\, \chi[\phi,\tau_{E}^{i}].
\end{equation}
Here T represents time ordering with respect to $\tau_{E}^{'}$.
  The crucial point
is that the Euclidean time evolution operator, $\hat{U}_{E}$, is non-unitary.
This is natural since we know that wavefunctions decay exponentially during
tunneling.  If $\hat{U}_{E}$ was unitary, the easiest way to calculate it
would be to transform to the Heisenberg picture, solve the field equations
mode by mode, and compute Bogolubov coefficients.  However, as emphasized in
Ref.~\cite{Rub84} the non-unitarity of $\hat{U}_{E}$ implies that the
Schr\"{o}dinger and Heisenberg pictures are inequivalent, making the standard
method inapplicable.  Instead, one can use the method developed in
Ref.~\cite{Rub84} which closely resembles the standard one but is more general.
 We first describe the state right before tunneling.  For convenience, set
$x_{m}^{i}=\tau^{i}=\tau_{E}^{i}=0$.  Let $\xi_{\omega}(x,\tau)$ be a complete
set of positive norm solutions to the Klein-Gordon equation which vanish
vanish at the mirror:
\begin{equation}
\left[-\frac{\partial^{2}}{\partial \tau^{2}} + \frac{\partial^{2}}
{\partial x^{2}}\right] \xi_{\omega}(x,\tau)=0
\end{equation}
\begin{equation}
i \int dx \left[\xi_{\omega}^{*}(x,\tau)\,\frac{\partial}{\partial \tau}
\xi_{\omega^{'}}(x,\tau)\,
-\frac{\partial}{\partial \tau}\xi_{\omega}^{*}(x,\tau)\xi_{\omega^{'}}
(x,\tau)\right]
=\delta_{\omega \omega^{'}}
\end{equation}
\begin{equation}
\xi_{\omega}(x_{m}(\tau),\tau)=0.
\end{equation}
The set of allowed frequencies $\omega$ is taken to be discrete, and
$\sum_{\omega}$ represents summation over this set. The field operators can
then be expanded in terms of these modes:
\begin{equation}
\hat{\phi}(x,\tau)=\sum_{\omega}\left[\hat{a}_{\omega} \xi_{\omega}(x,\tau)
+\hat{a}_{\omega}^{\dagger}\xi_{\omega}^{*}(x,\tau)\right]
\end{equation}
\begin{equation}
\hat{\pi}_{\phi}(x,\tau)=\frac{\partial}{\partial \tau}\hat{\phi}(x,\tau)
=\sum_{\omega}\left[
\hat{a}_{\omega}\frac{\partial}{\partial \tau}
\xi_{\omega}(x,\tau)+\hat{a}_{\omega}^{\dagger}
\frac{\partial}{\partial \tau}\xi_{\omega}(x,\tau)\right]
\end{equation}
with $\left[\hat{a}_{\omega},\hat{a}_{\omega'}^{\dagger}\right]=\delta_{
\omega \omega '}$.

 Now define Euclidean fields $\hat{\phi}^{E}(y,\tau_{E})$,
$\hat{\pi}_{\phi}^{E}(y,\tau_{E})$ which agree with $\hat{\phi}(x,\tau)$,
$\hat{\pi}_{\phi}(y,\tau)$ at $\tau=\tau^{E}=0$, but evolve according to
\begin{equation}
\hat{\phi}^{E}(y,\tau_{E})=\hat{U}_{E}^{-1}(\tau_{E},0)\,\hat{\phi}^{E}(y,0)\,
\hat{U}_{E}(\tau_{E},0)
\label{e:phi}
\end{equation}
\begin{equation}
\hat{\pi}_{\phi}^{E}(y,\tau_{E})=\hat{U}_{E}^{-1}(\tau_{E},0)\,
\hat{\pi}_{\phi}^{E}(y,0)\, \hat{U}_{E}(\tau_{E},0).
\label{e:pi}
\end{equation}
We will calculate $\hat{U}_{E}(\tau_{E},0)$ by first finding $\hat{\phi}
^{E}(y,\tau_{E})$, $\hat{\pi}^{E}_{\phi}(y,\tau_{E})$.
 The field equations for these operators are
\begin{equation}
\frac{\partial \hat{\phi}^{E}}{\partial \tau_{E}} = -\left[\hat{\phi}^{E},
\hat{H}_{\phi}^{E}\right]=-i\hat{\pi}_{\phi}^{E}+\frac{dx_{m}}{d\tau_{E}}
\frac{\partial \hat{\phi}^{E}}{\partial y}
\end{equation}
\begin{equation}
\frac{\partial \hat{\pi}_{\phi}^{E}}{\partial \tau_{E}}=
-\left[\hat{\pi}_{\phi}^{E},\hat{H}_{\phi}^{E}\right]=
-i\frac{\partial^{2} \hat{\phi}^{E}}{\partial y^{2}}
+\frac{dx_{m}}{d\tau_{E}}\frac{\partial \hat{\pi}_{\phi}^{E}}{\partial y}.
\end{equation}
So
\begin{equation}
\hat{\pi}_{\phi}^{E}=i\left(\frac{\partial \hat{\phi}^{E}}{\partial \tau_{E}}
-\frac{dx_{m}}{d\tau_{E}}\frac{\partial \hat{\phi}^{E}}{\partial y}\right)
\end{equation}
and
\begin{equation}
\frac{\partial^{2} \hat{\phi}^{E}}{\partial \tau_{E}^{\;2}}
+\left[1+\left(\frac{dx_{m}}{d\tau_{E}}\right)^{2}\right]\frac{\partial^{2}
\hat{\phi}^{E}}{\partial y^{2}} - 2\frac{dx_{m}}{d\tau_{E}}
\frac{\partial^{2} \hat{\phi}^{E}}{\partial y \partial \tau_{E}}-\frac
{d^{2}x_{m}}{d\tau_{E}^{\;2}}\frac{\partial \hat{\phi}^{E}}{\partial y}=0.
\label{e;phi}
\end{equation}
Equation (\ref{e;phi}) can be obtained by varying the action
\begin{equation}
S=\frac{1}{2}\int dy\, d\tau_{E}\, \sqrt{g_{E}\,}\, g_{E}^{\mu \nu}
\partial_{\mu}\phi\partial_{\nu}\phi
\end{equation}
with the Euclidean metric
\begin{equation}
ds_{E}^{2}=g^{E}_{\mu \nu}dx^{\mu}dx^{\nu}=d\tau_{E}^{\;2}+2\,\frac{dx_{m}
}{d\tau_{E}}\,dx\,d\tau_{E}+dx^{\;2}.
\end{equation}
$\hat{\phi}^{E}$, $\hat{\pi}_{\phi}^{E}$ can be expanded in terms of modes
$f_{\omega}$ which satisfy the Euclidean Klein-Gordon equation (\ref{e;phi})
and which vanish at $y=0$,
\begin{equation}
\hat{\phi}^{E}(y,\tau_{E})=\sum_{\omega}\hat{b}_{\omega}f_{\omega}(y,
\tau_{E})
\label{e:sta}
\end{equation}
\begin{equation}
\hat{\pi}_{\phi}^{E}(y,\tau_{E})=i\sum_{\omega}\hat{b}_{\omega}
\left(\frac{\partial}{\partial \tau_{E}}f_{\omega}(y,\tau_{E})
-\frac{dx_{m}}{d\tau_{E}}\frac{\partial}{\partial y}f_{\omega}(y,\tau_{E})
\right).
\end{equation}

 As the Euclidean Klein-Gordon equation is elliptic, one cannot in general
impose Cauchy boundary conditions at $\tau_{E}=0$  on $f_{\omega}$.  The
resulting solutions would not satisfy the mirror boundary condition.  With the
appropriate boundary conditions, either Dirichlet or Neumann, imposed at
$\tau_{E}=0$ and $\tau_{E}=\tau_{E}^{f}$, a detailed calculation is, of
course, required to find $f_{\omega}$ for a generic mirror trajectory. We shall
take the solutions as given and only use their specific forms in a region far
from the mirror, where they are simple.

Now, using the condition that the two sets of operators $\hat{\phi}$,
 $\hat{\pi}_{\phi}$ and $\hat{\phi}^{E}$, $\hat{\pi}_{\phi}^{E}$ are equal at
$\tau=\tau_{E}=0$, and taking inner products, the operators $\hat{b}_{\omega}$
can be expressed as a linear combination of $\hat{a}_{\omega}$,
$\hat{a}_{\omega}^{\dagger}$:
\begin{equation}
\hat{b}_{\omega}=\sum_{\omega'}\left[\alpha_{\omega \omega'}\hat{a}_{\omega'}
+\beta_{\omega \omega'}\hat{a}_{\omega'}^{\dagger}\right].
\end{equation}
Then using
\begin{equation}
\hat{\phi}^{E}(y,\tau_{E}^{f})=\hat{U}_{E}^{-1}(\tau_{E}^{f},0)\,
\hat{\phi}_{E}(y,0)\,\hat{U}_{E}(\tau_{E}^{f},0)=\hat{U}_{E}^{-1}
(\tau_{E}^{f},0)\,
\hat{\phi}(y,0)\,\hat{U}_{E}(\tau_{E}^{f},0)
\end{equation}
and the analogous expression for $\hat{\pi}_{\phi}^{E}$, the following
equations for $\hat{U}^{E}$ are obtained:
\pagebreak
$$
\sum_{\omega}\sum_{\omega'}\left[\alpha_{\omega \omega'} \hat{a}_{\omega'}
+\beta_{\omega \omega'} \hat{a}_{\omega'}^{\dagger}\right]
f_{\omega}(y,\tau_{E}^{f})
$$
\begin{equation}
=\sum_{\omega}\left[\hat{U}_{E}^{-1}(\tau_{E}^{f},0)\,\hat{a}_{\omega}\,
\hat{U}_{E}(\tau_{E}^{f},0)\,\xi_{\omega}(y,0)
+\hat{U}_{E}^{-1}(\tau_{E}^{f},0)\,\hat{a}_{\omega}^{\dagger}\,
\hat{U}_{E}(\tau_{E}^{f},0)\,\xi_{\omega}^{*}(y,0)\right]
\end{equation}
and
$$
i\sum_{\omega}\sum_{\omega'}\left[\alpha_{\omega \omega'}\hat{a}_{\omega'}
+\beta_{\omega \omega'}\hat{a}_{\omega'}^{\dagger}\right]
\frac{\partial}{\partial \tau_{E}}f_{\omega}(y,\tau_{E}^{f})
$$
\begin{equation}
=\sum_{\omega}\left[\hat{U}_{E}^{-1}(\tau_{E}^{f},0)\,\hat{a}_{\omega}\,
\hat{U}_{E}(\tau_{E}^{f},0)\,\frac{\partial}{\partial \tau}\xi_{\omega}(y,0)
+\hat{U}_{E}^{-1}(\tau_{E}^{f},0)\,\hat{a}_{\omega}^{\dagger}\,
\hat{U}_{E}(\tau_{E}^{f},0)\,\frac{\partial}{\partial \tau}
\xi_{\omega}^{*}(y,0)
\right].
\end{equation}
Again taking inner products, this leads to relations of the form
\begin{equation}
\hat{U}_{E}^{-1}(\tau_{E}^{f},0)\,\hat{a}_{\omega}\,\hat{U}_{E}(\tau_{E}^{f},0)
=\sum_{\omega'}\left[u_{\omega \omega'}\hat{a}_{\omega'}
+v_{\omega \omega'}\hat{a}_{\omega'}^{\dagger}\right]
\label{e:tev}
\end{equation}
\begin{equation}
\hat{U}_{E}^{-1}(\tau_{E}^{f},0)\,\hat{a}_{\omega}^{\dagger}\,
\hat{U}_{E}(\tau_{E}^{f},0)=\sum_{\omega'}\left[w_{\omega \omega'}
\hat{a}_{\omega'}+z_{\omega \omega'}\hat{a}_{\omega'}\right].
\end{equation}
Then it can be shown that \cite{Rub84}
\begin{equation}
\hat{U}_{E}(\tau_{E}^{f},0)=\mbox{ const. }\times:\exp\sum_{\omega}
\sum_{\omega'}\left[\frac{1}{2}D_{\omega \omega'}\hat{a}_{\omega}^{\dagger}
\hat{a}_{\omega'}^{\dagger}+F_{\omega \omega'}\hat{a}_{\omega}\hat{a}
_{\omega'}+\frac{1}{2}G_{\omega \omega'}\hat{a}_{\omega}\hat{a}_{\omega'}
\right]:
\end{equation}
where the matrices $D$, $F$, and $G$ are defined by
\begin{equation}
D=vz^{-1}\mbox{ \ ; \ } F=\left(z^{T}\right)^{-1}-1\mbox{ \ ; \ }G=-z^{-1}w.
\end{equation}
The state after tunneling is then determined,
\begin{equation}
\left|\chi(\tau_{E}^{f})\right\rangle = \hat{U}_{E}(\tau_{E}^{f})
\left|\chi(0)\right\rangle
\label{end}
\end{equation}
and is expressed in terms of occupation numbers with respect to the modes
$\xi_{\omega}(y,0)$, where now $y=x-x_{m}^{f}$.  All of the information about
the final state is contained in the matrices $D$, $F$, and $G$, which are in
turn given in terms of inner products between the modes $f_{\omega}$ and
$\xi_{\omega}$.

As a simple application of these formul{\ae} we will calculate the shift in the
spectrum of outgoing particles which are far from the mirror at the time of
tunneling. It is assumed that the mirror was initially at rest and the field
in its ground state. The mirror subsequently accelerates in the potential
$V(x_{m})$ until it reaches the classical turning point $x_{m}^{i}$. It is well
known that as a result of the mirror's acceleration, a flux of outgoing
particles is created whose spectrum is calculable by standard methods
\cite{Dav77}.  Outgoing particles far from the mirror are wavepackets composed
of superpositions of plane waves,
\begin{equation}
\xi_{\omega}(x,\tau)=\frac{1}{2\sqrt{\omega}}\,e^{-i\omega(\tau-x)}
\end{equation}
The spectrum of outgoing particles located at $x=\bar{x}\gg\omega^{-1}$
 at $\tau=0$ is written as
\begin{equation}
\sum_{\{n_{\omega}\}}S_{\bar{x}}\left(\{n_{\omega}\}\right)\left|\{n_{\omega}\}
\right\rangle
\end{equation}
where $\{n_{\omega}\}$ is a set of occupation numbers and $S_{\bar{x}}\left(
\{n_{\omega}\}\right)$ is the amplitude for the set to occur.

Far from the mirror, the modes $f_{\omega}$ are easy to calculate since the
mirror boundary condition is irrelevant. They are of two types,
$$
f_{\omega}^{\mbox{--}}=\frac{1}{2\sqrt{\omega}}\,e^{-\omega \tau_{E}+i\omega x}
=\frac{1}{2\sqrt{\omega}}\,e^{-\omega \tau_{E}+i\omega \left(y+x_{m}(\tau_{E}
)\right)}
$$
\begin{equation}
f_{\omega}^{\mbox{+}}=\frac{1}{2\sqrt{\omega}}\,e^{\omega \tau_{E}+i\omega x}
=\frac{1}{2\sqrt{\omega}}\,e^{\omega \tau_{E}+i\omega \left(
y+x_{m}(\tau_{E})\right)}
\end{equation}
Then $\hat{\phi}$, $\hat{\pi}$ and $\hat{\phi}^{E}$, $\hat{\pi}_{\phi}^{E}$
are equal at $\tau=\tau_{E}=0$ if
\begin{equation}
\hat{b}_{\omega}^{\mbox{--}}=\hat{a}_{\omega} \mbox{ \ ; \ }
 \hat{b}_{\omega}^{\mbox{+}}=\hat{a}_{\omega}^{\dagger}.
\end{equation}
Equation (\ref{e:tev}) gives:
$$
\hat{U}_{E}^{-1}(\tau_{E}^{f},0)\,\hat{a}_{\omega}\,\hat{U}_{E}(\tau_{E}
^{f}),0)
=e^{-\omega \tau_{E}^{f}+i\omega x_{m}^{f}}\,\hat{a}_{\omega}
$$
\begin{equation}
\hat{U}_{E}^{-1}(\tau_{E}^{f},0)\,\hat{a}_{\omega}^{\dagger}\,
\hat{U}_{E}(\tau_{E}^{f},0)=e^{\omega \tau_{E}^{f}+i\omega x_{m}^{f}}
\,\hat{a}_{\omega}^{\dagger}
\end{equation}
leading to
\begin{equation}
D=G=0 \mbox{ \ ; \ } F_{\omega \omega'}=\left(e^{-\omega \tau_{E}^{f}
-i\omega x_{m}^{f}}-1\right)\delta_{\omega \omega'}
\end{equation}
and
$$
\hat{U}_{E}(\tau_{E}^{f},0)=\mbox{ const. }\times :\exp\sum_{\omega}
\left[e^{-\omega \tau_{E}^{f}-i\omega x_{m}^{f}}-1\right]\hat{a}_{\omega}
^{\dagger}\hat{a}_{\omega}:
$$
\begin{equation}
=\mbox{ const. }\times:\exp\sum_{\omega}\left[e^{-i\omega x_{m}^{f}}
-1\right]\hat{a}_{\omega}^{\dagger}\hat{a}_{\omega}::\exp
\sum_{\omega}\left[e^{-\omega \tau_{E}^{f}}-1\right]\hat{a}_{\omega}^{\dagger}
\hat{a}_{\omega}:
\end{equation}
The first factor is a translation operator which expresses the state in terms
of the modes $\xi_{\omega}(x,0)$ instead of $\xi_{\omega}(x+x_{m}^{f},0)$, and
the second factor acts on a state $\left|\{n_{\omega}\}\right\rangle$ to give
$e^{-E\left(\{n_{\omega}\}\right)\tau_{E}^{f}}\left|\{n_{\omega}\}
\right\rangle$, where $E\left(\{n_{\omega}\}\right)=\sum n_{\omega}\omega$
 is the energy of the state. Therefore, the state after tunneling is
\begin{equation}
\mbox{const. }\times\sum_{\{n_{\omega}\}}e^{-E\left(\{n_{\omega}\}\right)
\tau_{E}^{f}}\,S_{\bar{x}}\left(\{n_{\omega}\}\right)\left|\{n_{\omega}\}
\right\rangle.
\label{e:spec}
\end{equation}
The result of the tunneling is simply to shift the spectrum from $S_{\bar{x}}$
to $e^{-E({n_{\omega}})\tau_{E}^{f}}S_{\bar{x}}$.

It is not difficult to understand this result.  Since the total energy is fixed
, the state before tunneling is given by a superpostion of the various ways
of distributing the energy between the mirror and the radiation.  As the
mirror's probability to tunnel depends exponentially on its energy, we expect
an inverse exponential correlation  between tunneling and energy in radiation.
Thus an observer measuring the spectrum of radiation, conditional on the mirror
tunneling, finds the result (\ref{e:spec}).
 Far from the mirror the shift in the
spectrum depends only on $\tau_{E}^{f}$, the amount of Euclidean time spent
during tunneling.  This is because the tunneling amplitude in the WKB
approximation is $e^{-S}$, and the derivative of $S$ with respect to energy is
just the Euclidean time.

If we were to identify the Euclidean time with an inverse temperature, the
shift would become a Boltzmann factor.  This makes it easy to generate thermal
distributions of radiation.  Specifically, if the distribution before tunneling
was a constant, then after tunneling tracing over the states of the mirror
would yield a thermal density matrix for the radiation. A number of authors
have been led by this fact to seek a connection between the thermal radiation
that arises in the contexts of cosmology and black holes and an occurrence of
tunneling \cite{Kan89,Bro91,Cas92}.
 Such a connection relies upon assumptions about what is on the
other side of the barrier and what the spectrum of radiation is there. In this
work we only consider situations
 where there is a well defined classical trajectory
on either side of the barrier;  we are interested in the case in which there is
collapsing matter on side of the barrier and a black hole on the other.  The
treatment of this process requires an extension of the previous method to
include gravity.

\newsection{Application to Gravity}

\label{secgrav}
In this section we make a WKB approximation to gravity in a manner which
directly parallels that for the moving mirror.  The starting point for the
 canonical quantization of gravity is to write the metric as
\begin{equation}
ds^{2}=-\left(N^{t}dt\right)^{2}+h_{ij}\left(dx^{i}+N^{i}dt\right)\left(
dx^{j}+N^{j}dt\right).
\end{equation}
With this definition, the action for gravity plus matter takes the form
$$
S=\frac{m_{p}^{\,2}}{16\pi}\int d^{4}x\sqrt{-g}\left({\cal R}-2\Lambda\right)
+S_{M}+\mbox{ boundary terms}
$$
\begin{equation}
=\int d^{4}x\left(\pi_{\phi_{i}}\dot{\phi}^{i}+\pi_{ij}\dot{h}^{ij}
-N^{t}{\cal H}_{t}-N_{i}{\cal H}^{i}\right)+\mbox{ boundary terms}
\end{equation}
with
$$
{\cal H}_{t}=\frac{8\pi}{m_{p}^{\,2}}h^{-\frac{1}{2}}\left(h_{ik}h_{jl}
+h_{il}h_{jk}-h_{ij}h_{kl}\right)\pi^{ij}\pi^{kl}
-\frac{m_{p}^{\,2}}{16\pi}h^{\frac{1}{2}}\left(^{3}{\cal R}-2\Lambda\right)+
{\cal H}_{t_{M}}
$$
\begin{equation}
=\frac{16\pi}{m_{p}^{\,2}}G_{ijkl}\pi^{ij}\pi^{kl}-\frac{m_{p}^{\,2}}{16\pi}
h^{\frac{1}{2}}\left(^{3}{\cal R}-2\Lambda\right)+{\cal H}_{t_{M}}
\end{equation}
\begin{equation}
{\cal H}^{i}=-2\pi^{ij}_{\;\; | j}+{\cal H}_{M}^{i}
\end{equation}
Covariant differentiation and the raising and lowering of indices are performed
with respect to the spatial metric $h_{ij}$. $\phi_{i}$ refer to arbitrary
matter fields.

The boundary term in the action is determined by requiring that no such term
arise in the variation of the Hamiltonian \cite{Reg74}. Restricting to
 asymptotically flat metrics with $N^{t}\rightarrow 1$, $N_{i}\rightarrow 0$
 as $r\rightarrow\infty$, but allowing for time translations at spatial
 infinity, the appropriate boundary term is numerically equal to the mass of
 the system. Thus the Hamiltonian is
\begin{equation}
H=\int d^{3}x \left[N^{t}{\cal H}_{t}+N_{i}{\cal H}^{i}\right] + M.
\end{equation}
Varying the action with respect to $N^{t}$ and $N_{i}$ yields the constraints
\begin{equation}
{\cal H}_{t}={\cal H}^{i}=0.
\end{equation}
The system is quantized by making the replacements
\begin{equation}
\pi_{ij}\rightarrow-i\frac{\delta}{\delta h_{ij}} \mbox{ \ ; \ }
\pi_{\phi_{i}}\rightarrow-i\frac{\delta}{\delta\phi_{i}}
\end{equation}
and demanding that the state satisfy the constraints $\hat{{\cal H}}_{t}\Psi
=\hat{{\cal H}}^{i}\Psi=0$ as well as the Schr\"{o}dinger equation
$H\Psi=i\partial \Psi/\partial t$.

The constraints
\begin{equation}
\hat{{\cal H}}^{i}\Psi=\left[2i\left(\frac{\delta}{\delta h_{ij}}\right)
_{|j}+\hat{{\cal H}}_{M}^{i}\right]\Psi=0
\end{equation}
enforce invariance of the state under spatial reparameterizations, and the
constraint
\begin{equation}
\hat{{\cal H}}_{t}\Psi=\left[-\frac{16\pi}{m_{p}^{\;2}}G_{ijkl}
\frac{\delta}{\delta h_{ij}}\frac{\delta}{\delta h_{kl}}-\frac{m_{p}^{\;2}}
{16\pi}h^{\frac{1}{2}}\left(^{3}{\cal R}-2\Lambda\right)+\hat{{\cal H}}_{t_{M}}
\right]\Psi=0
\end{equation}
is the Wheeler-DeWitt equation. Proceeding as before, we seek a semiclassical
solution of the form
\begin{equation}
\Psi\left[h_{ij},\phi_{i}\right]=\psi_{VV}[h_{ij}]\,e^{im_{p}^{\;2}S[h_{ij}]}
\,\chi\left[\phi_{i},h_{ij}\right].
\end{equation}
At first order the Einstein-Hamilton-Jacobi equation is obtained:
\begin{equation}
\frac{16\pi}{m_{p}^{\;2}}G_{ijkl}\frac{\delta S}{\delta h_{ij}}\frac{\delta S}
{\delta h_{kl}}-\frac{m_{p}^{2\;}}{16\pi}h^{\frac{1}{2}}\left(^{3}{\cal R}-
2\Lambda \right)=0.
\label{e:ehj}
\end{equation}
Zeroth order yields
\begin{equation}
-\frac{16\pi}{m_{p}^{\;2}}i\,G_{ijkl}\frac{\delta S}{\delta h_{ij}}\frac{\delta
\chi}{\delta h_{kl}}+\hat{{\cal H}}_{t_{M}}\chi=0
\label{e:sc}
\end{equation}
provided $\psi_{VV}$ satisfies
\begin{equation}
G_{ijkl}\frac{\delta^{2}S}{\delta h_{ij} \delta h_{ij}}\psi_{VV}
+G_{ijkl}\frac{\delta S}{\delta h_{ij}}\frac{\delta \psi_{VV}}{\delta h_{kl}}
=0.
\end{equation}
The momentum constraints at first order are
\begin{equation}
\left(\frac{\delta S}{\delta h_{ij}}\right)_{|j}=0
\end{equation}
and at zeroth order are
\begin{equation}
2i\left(\frac{\delta \chi}{\delta h_{ij}}\right)_{|j}+\hat{{\cal H}}_
{M}^{i}\,\chi=0.
\label{e:mo}
\end{equation}

Equations (\ref{e:sc}) and (\ref{e:mo}) describe how the matter wave function
evolves as the spatial geometry changes. Quantum field theory in curved space
can be recovered by writing $\chi$'s dependence on $h_{ij}$ in terms of a
time functional $\tau[x;h_{ij}]$, and by reintroducing a lapse $N^{\tau}$ and
 shift $N_{i}$, demanding that they obey
\begin{equation}
G_{ijkl}\frac{\delta S}{\delta h_{ij}}=\frac{m_{p}^{\,2}}{16\pi N^{\tau}}
\left(\int dy \frac{\delta h_{kl}}{\delta \tau[y;h_{ab}]}
-N_{i|j}-N_{j|i}\right).
\label{e:con}
\end{equation}
Then
\begin{equation}
-i\frac{16\pi}{m_{p}^{\;2}}\int\left(N^{\tau}G_{ijkl}\frac{\delta S}
{\delta h_{ij}} \frac{\delta \chi}{\delta h_{kl}}+2iN_{i}\left(\frac{\delta
 \chi}{\delta h_{ij}}\right)_{|j}\right)=-i\int\frac{\delta h_{ij}}{\delta
 \tau}\frac{\delta \chi}{\delta h_{ij}}
\end{equation}
so that the equation for $\chi$ becomes
\begin{equation}
\int d^{3}x
\left[N^{\tau}\hat{{\cal H}}_{t_{M}}+N_{i}\hat{{\cal H}}_{M}^{i}\right]
\chi[\phi_{i};\tau]= i\frac{\partial}{\partial \tau}\chi[\phi_{i};\tau].
\label{e:scg}
\end{equation}
The condition (\ref{e:con}) agrees with the classical relation between
$\pi_{ij}$ and $h_{ij}$, demonstrating that $\tau[x;h_{ij}]$ is the classical
time and that (\ref{e:scg}) is the Schr\"{o}dinger picture version of quantum
field theory in curved space.

As with the mirror example, $\tau$ becomes imaginary during tunneling so we
define a Euclidean time $\tau_{E}$ along with a Euclidean lapse $N^{\tau_{E}}
=iN^{\tau}$, in terms of which $\chi$ obeys
\begin{equation}
-\int dx [N^{\tau_{E}}\hat{{\cal H}}_{t_{M}}+iN_{i}\hat{{\cal H}}_{M}^{i}]\chi
[\phi_{i},\tau_{E}]=\frac{\partial}{\partial \tau_{E}}\chi[\phi_{i},\tau_{E}].
\end{equation}
For a massless scalar field with action
\begin{equation}
S=-\frac{1}{2} \int d^{4}x \sqrt{-g\,}g^{\mu \nu}\,\partial_{\mu}\phi
 \partial_{\nu}\phi,
\end{equation}
we have
\begin{equation}
\hat{{\cal H}}_{t_{M}}=\frac{1}{2}\left(h^{-\frac{1}{2}}\hat{\pi}_{\phi}^{2}
+h^{\frac{1}{2}} h^{ij} \partial_{i}\hat{\phi}\, \partial_{j} \hat{\phi}\right)
\end{equation}
\begin{equation}
\hat{{\cal H}}_{i_{M}}=\partial_{i}\hat{\phi}\, \hat{\pi}_{\phi}.
\end{equation}
To evolve $\chi$ through the tunneling region one is required to calculate the
Euclidean time evolution operator
\begin{equation}
\hat{U}_{E}(\tau_{E}^{f},\tau_{E}^{i})=T \exp\left[-\int_{\tau_{E}^{i}}
^{\tau_{E}^{f}} \hat{{\cal H}}^{E}_{\phi} d\tau_{E}\right]
\end{equation}
with
\begin{equation}
\hat{H}^{E}_{\phi}=\int d^{3}x\, [N^{\tau_{E}}(-\frac{1}{2}h^{-\frac{1}{2}}
\frac{\delta^{2}}{\delta \phi^{2}} +\frac{1}{2}h^{\frac{1}{2}}h^{ij}
\partial_{i}\phi\, \partial_{j} \phi)+N_{i} \partial_{i} \phi \frac{\delta}
{\delta \phi}].
\end{equation}
As before, one proceeds by defining Euclidean fields obeying
(\ref{e:phi},\ref{e:pi}). In the present case the resulting field equations
are:
$$
\left(\sqrt{g_{E}}\,g_{E}^{\mu \nu}\partial_{\mu}\hat{\phi}^{E}\right)_{,\nu}=0
$$
\begin{equation}
\hat{\pi}_{\phi}^{E}=i\frac{h^{\frac{1}{2}}}{N^{\tau_{E}}}\left(
\frac{\partial \hat{\phi}_{E}}{\partial \tau_{E}}-N^{i}\partial_{i}\phi\right)
\end{equation}
with
\begin{equation}
ds_{E}^{\;2}=g_{\mu \nu}^{E}\,dx^{\mu} dx^{\nu}=
\left(N^{\tau_{E}}d\tau_{E}\right)^{2}
+h_{ij}\left(dx^{i}+N_{i}d\tau_{E}\right)\!\left(dx^{j}+N^{j}d\tau_{E}\right).
\end{equation}
The evolution operator, and therefore the state after tunneling, is determined
by solving the field equations mode by mode, and repeating the steps leading
from (\ref{e:sta}) to (\ref{end}).
\label{s:grav}

\newsection{Black Hole Radiation in the Presence of Tunneling}

\label{sechole}

We can now apply this method to determine how the radiation from a black hole
is affected by tunneling. It is well known that a black hole formed
classically from collapsing matter radiates in a complicated manner at early
times due to the time dependent geometry, but at late times will inevitably
radiate as a black body at the Hawking temperature. Is this scenario altered
if the black hole is formed while tunneling?  We shall show that it is not.
The form of the late time radiation is insensitive to the hole's
unconventional history in a way that is consistent with the intuitive picture
of Hawking radiation being caused by pair production near the horizon.

We consider the behaviour of a scalar field on the background of a false
vacuum bubble which tunnels leading to the formation of a black hole.  The
action for a false vacuum bubble in the thin wall approximation is
\begin{equation}
S=\frac{1}{16\pi} \int d^{4}x \sqrt{-g\,}\,{\cal R} -\frac{\Lambda_{I}}
{8\pi}\int_{bubble} \! d^{4}x \sqrt{-g\,}\, -\frac{\mu}{4\pi}\int_{wall}d^{3}A
\end{equation}
where $\Lambda_{I}$ is the cosmological constant of the false vacuum, and
$\mu$ is the energy density of the bubble wall.  The classical solutions for
this action have been derived in Refs.~\cite{Sato81}-\cite{Guth87}.
  In what follows
we refer to the treatment of Ref.~\cite{Guth87}.  The spherically symmetric
solutions are characterized by three parameters: $\Lambda_{I}$, $\mu$, and
the total mass $M$.  In addition, for given $\Lambda_{I}$ and $\mu$ there is
a critical mass $M_{cr}$ below which there are two solutions: type (a),
where the bubble emerges from a singularity with zero  radius, subsequently
expands to a maximum radius, and then recollapses; type (b), where the bubble
initially collapses from infinite radius, reaches a minimum radius, and then
reexpands.  Using the results of Refs.~\cite{Guth90,Pol90}, we focus on an
expanding solution of type (a) which tunnels to an expanding solution of
type (b).
 We confine our interest to the region outside the bubble where
the metric, written in terms of Schwarzschild time $t$ and $r_{*}=r+2M
\ln({r}/{2M}-1)$, is
\begin{equation}
ds^{2}=\left(1-\frac{2M}{r}\right)\left(-dt^{2}+dr_{*}^{2}\right)+r^{2}
d\Omega^{2}.
\end{equation}
As $t$ and $r_{*}$ cover only part of the complete manifold, we introduce
Kruskal-Szekeres coordinates,
\begin{equation}
ds^{2}=\frac{32M^{3}e^{-r/2M}}{r}(-dT^{2}+dX^{2}) +r^{2}d\Omega^{2}.
\end{equation}
The two sets of coordinates are related by
$$
\left(\frac{r}{2M}-1\right)e^{r/2M}=X^{2}-T^{2}
$$
\begin{equation}
t=\left\{\begin{array}{ll}
4M\tanh^{-1}(T/X) & \mbox{if \, $|T/X|<1$}\\
4M\tanh^{-1}(X/T) & \mbox{if \, $|T/X|>1$}
\end{array}\right.
\end{equation}
Using these cordinates the type (a) and (b) solutions of interest are depicted
in Fig. 2.

The tunneling amplitude for this process has been computed by two different
methods.  In Ref.~\cite{Pol90} the solution to the Wheeler-DeWitt equation
is found in the
WKB approximation by solving the Einstein-Hamilton-Jacobi equation
 (\ref{e:ehj}). Since the solution behaves as $e^{-S}$, and the tunneling
amplitude is given by the ratio of the wavefunction evaluated at the initial
and final geometries, the tunneling amplitude is
\begin{equation}
\exp\left(S[h_{ij}^{\mbox{\scriptsize{initial}}}]-S[h_{ij}^
{\mbox{\scriptsize{final}}}]\right)
\end{equation}
No difficulties arise in this approach; the calculation of tunneling
amplitude proceeds in a straightforward fashion.

In Ref.~\cite{Guth90} the calculation is performed using the functional
 integral.
In this formalism one looks for a manifold which interpolates between the
initial and final surfaces and which is a solution to the Euclidean Einstein
equations. The tunneling amplitude is $e^{-S}$, where S is the action of the
solution.  It is found, however, that solving the field equations leads to a
sequence of three geometries which do not form a manifold.  To see this, first
note that the geometry outside the bubble is Euclidean Schwarzschild space,
obtained by $t\rightarrow it_{E}$, $T\rightarrow iT_{E}$,
\begin{equation}
ds_{E}^{2}=\left(1-\frac{2M}{r}\right)\left(dt_{E}^{2}+dr_{*}^{2}\right)
+r^{2}d\Omega^{2}
=\frac{32M^{3}e^{-r/2M}}{r}\left(dT^{2}+dX^{2}\right)+r^{2}d\Omega^{2}
\end{equation}
with
\begin{equation}
\left(\frac{r}{2M}-1\right)e^{r/2m}=X^{2}+T_{E}^{2} \mbox{\ ;\ }
t_{E}=4M\tan^{-1}(T_{E}/{X}).
\end{equation}
It remains to describe the motion of the bubble wall.  Solving the equations
of motion leads to the trajectory in Fig. 3. It is seen that the bubble wall
crosses the initial surface during the course of its motion, creating a
situation in which it is impossible to identify a region which is swept out
by the evolving hypersurface.  Some regions of the manifold are crossed
twice by the hypersurface, some once, and some not at all.  The authors of
Ref.~\cite{Guth90} call
this object a pseudomanifold and give a prescription to calculate its action
by assigning covering numbers to the various regions, but this is not needed
for what follows.

With these results in hand, the technique of Sect.~(3) can be
used to calculate the state of the scalar field after tunneling. It was seen
that once the solution of the Einstein-Hamilton-Jacobi equation is given, the
field wave functional $\chi$ is fully determined by equations (\ref{e:sc}) and
(\ref{e:mo}).  Since $S[h_{ij}]$ is calculated in Ref.~\cite{Pol90},
 we have all
 that we need to find $\chi$.   This would, however, require finding the
 solution
to an unfamiliar functional differential equation. To cast it in in the form
of the Schr\"{o}dinger equation a lapse $N^{\tau_{E}}$, shift $N_{i}$
 and time $\tau_{E}$ were reintroduced leading to the appearance of the
Euclidean metric $g_{\mu \nu}^{E}$.  In the present case there is no true
interpolating Euclidean manifold, so that any choice of $N^{\tau_{E}}$ and
$N_{i}$ which define a well behaved $g_{\mu \nu}^{E}$ will lead to a bubble
trajectory that is a multivalued function of time.  Alternatively, a choice
of time functional which gives a single valued bubble trajectory will
necessarily lead to a Euclidean metric with vanishing determinant at some
point.  In either case, it is not clear that the resulting Schr\"{o}dinger
equation is well defined.  This is apparent from Fig. 3, where it can be seen
that boundary conditions imposed on the initial surface and on the bubble wall
may contradict each other.  These difficulties arise as a result of trying to
compute the final state of the field in one step, which requires a Euclidean
manifold interpolating all the way from the initial surface to the final
surface, and can be avoided by calculating the state on a series of
intermediate hypersurfaces.  In this approach, it does not matter that the
bubble wall eventually crosses the initial surface since once the state is
 calculated at some intermediate point we can forget about what preceded it.

For simplicity, we will consider only the s-wave component of the scalar field
and frequencies high enough such that the geometrical optics approximation is
valid.  This means that the field equation is taken to be
\begin{equation}
g_{E}^{\mu \nu}\partial_{\mu} \partial_{\nu} \phi =0.
\end{equation}
The state of the field on the initial surface, $t=T=0$, is most conveniently
expressed in terms of the coordinates $r_{*}$ and $t$.  We divide the modes
into ingoing and outgoing,
$$
\xi_{\omega}^{\mbox{\scriptsize{in}}}(r_{*},t)=C_{\omega}\,
e^{-i\omega(t+r_{*})}
$$
\begin{equation}
\xi_{\omega}^{\mbox{\scriptsize{out}}}(r_{*},t)=
C_{\omega}\,e^{-i\omega(t-r_{*})}
\end{equation}
and write the field operator as
\begin{equation}
\hat{\phi}(r_{*},t)=\sum_{\omega}\left[\hat{a}_{\omega}^
{\mbox{\scriptsize{in}}}
\xi_{\omega}^{\mbox{\scriptsize{in}}}+\hat{a}_{\omega}^
{\mbox{\scriptsize{in}}\dagger}\xi_{\omega}^{\mbox{\scriptsize{in}*}}
(r_{*},t) + \mbox{ in} \rightarrow \mbox{out}\right].
\end{equation}
$C_{\omega}$ are normalization constants whose values will not be important.
We shall only consider the {\em in} modes as the treatment of the {\em out}
 modes is exactly the same. We also suppress the {\em in} superscript.

In the first stage of the evolution the hypersurface is pivoted around
$r_{*}=r_{*}^{b}$ by $180^{\circ}$, where $r_{*}^{b}$ is the position
 of the bubble wall on the
initial surface. The solutions to the Euclidean field
equations are most conveniently obtained by choosing Cauchy boundary conditions
on the initial surface, (clearly a valid procedure in this case)
$$
f_{\omega}^{+}(r_{*},0)=\xi_{\omega}(r_{*},0) \mbox{ \ ; \ }
\frac{\partial}{\partial t_{E}}f_{\omega}^{+}(r_{*},0)
=-i\frac{\partial}{\partial t}\xi_{\omega}^{*}(r_{*},0)
$$
\begin{equation}
f_{\omega}^{-}(r_{*},0)=\xi_{\omega}^{*}(r_{*},0) \mbox{ \ ; \ }
\frac{\partial}{\partial t_{E}}f_{\omega}^{-}(r_{*},0)
=-i\frac{\partial}{\partial t}\xi_{\omega}^{*}(r_{*},0)
\end{equation}
It is also easiest to use the $X$, $T$ coordinates as they are well behaved
everywhere.  Since the evolution of the hypersurface is simply a reflection
about the point $X=X^{b}$, a mode which has the form $f(X,T_{E})$ on the
 initial
surface has the form $f(-X+2X^{b},T)$ on the new surface. Using the relations
\begin{equation}
r_{*}=4M\ln\sqrt{X^{2}+T_{E}^{2}\,} \mbox{ \ ; \ } t_{E}=4M\tan^{-1}(T_{E}/X)
\end{equation}
and that
\begin{equation}
f_{\omega}^{\pm}(r_{*},t_{E})=C_{\omega}\,e^{\pm \omega t_{E} -i\omega r_{*}}
\end{equation}
near the initial surface, one sees that near the new surface,
\begin{equation}
f_{\omega}^{\pm}(X,T_{E})=C_{\omega}\exp\left(\frac{\mp 4M\omega
T_{E}}{-X+2X^{b}}
-4iM\omega\ln(-X+2X^{b})\right).
\end{equation}
Since on the new surface,
 $f_{\omega}^{+}=(f_{\omega}^{-})^{*}$ and $\partial f_
{\omega}^{+}/\partial t_{E}=-(\partial f_{\omega}^{-}/\partial t_{E})^{*}$,
 the
evolution operator $\hat{U}_{E}$ is unitary.  This means that the state on the
new surface has the same form as it did on the initial surface, but is now
expressed in terms of the modes
\begin{equation}
\xi_{\omega}(X,T)=C_{\omega}\exp\left(\frac{-4iM\omega T}{-X+2X^{b}}
+4iM\omega \ln(-X+2X^{b})\right).
\end{equation}
These modes can be approximated near $T=0$ as
\begin{equation}
\xi_{\omega}= \left\{\begin{array}{ll}
C_{\omega}\, e^{i\omega (t-r_{*})} & \mbox{ if $ \;  |X|\gg X^{b}$} \\
C_{\omega}\,e^{-(2iM\omega/X^{b})(T-X)} & \mbox{ if $ \;  |X|\ll X^{b}$}
\end{array} \right.
\end{equation}
Now it is useful to express the state in terms of modes which are nonzero only
inside or outside the horizon,
$$
\eta_{\omega}^{<}=\left\{\begin{array}{ll}
D_{\omega}\,e^{i\omega(t-r_{*})} & \mbox{ if $\; X<0$}\\
0 & \mbox{ if $ \; X>0$}
\end{array} \right.
$$
\begin{equation}
\eta_{\omega}^{>}=\left\{\begin{array}{ll}
0 & \mbox{ if $\; X<0$} \\
D_{\omega}\,e^{-i\omega(t+r_{*})} & \mbox{ if $\; X>0$.}
\end{array} \right.
\end{equation}
A fundamental result \cite{Haw75,Un76} in the derivation of black hole
radiance is that the vacuum state with respect to modes which have a time
dependence $e^{-i\omega T}$ is the state
\begin{equation}
\mbox{const.}\times \sum_{\{n_{\omega}\}}e^{-E(\{n_{\omega}\})/2T_{H}}
 \left| \{n_{\omega}\}\right\rangle_{<}\left| \{n_{\omega}\}\right\rangle_{>}
\label{e:dm}
\end{equation}
with respect to the modes $\eta_{\omega}^{<}$ and $\eta_{\omega}^{>}$.  The
sum runs over all sets of occupation numbers, $E=\sum n_{\omega}\omega$, and
$T_{H}=1/8\pi M$ is the Hawking temperature.  Further, near the horizon, any
deviation of $\left|\chi \right \rangle$ from the
 vacuum state can be ignored because
of the arbitrarily large redshift as $r_{*}\rightarrow -\infty$.  Far from the
horizon $\xi_{\omega}$ and $\eta_{\omega}^{<}$ agree so the form of the state
is unchanged there.

Now the hypersurface can be evolved the remainder of the way.  If we restrict
our attention to the region $X<X^{b}$, then the motion of the hypersurface
 is simply a translation, $t_{E}\rightarrow t_{E}-\Delta t_{E}$.
 This causes states with time dependence $e^{i\omega t}$ to be damped by a
 factor $e^{-\omega \Delta t_{E}}$, and states with time dependence
$e^{-i\omega t}$ to be amplified by a factor $e^{\omega \Delta t_{E}}$.
  Near the
horizon, the state $\left|\chi \right\rangle$ consists of pairs of positive and
negative frequency states according to (\ref{e:dm}).  One member of the pair
is damped but the other is amplified by a compensating amount so as to leave
the state $\left|\chi \right\rangle$ unchanged.
The final state of the field can then be summarized as follows.
  Far from the hole, where there is no pairing, the initial state is damped:
\begin{equation}
\sum_{\{n_{\omega}\}} S(\{n_{\omega}\})\left|\{n_{\omega}\}\right\rangle
\longrightarrow \mbox{const.}\times \sum_{\{n_{\omega}\}}e^{-E(\{n_{\omega}\})
\Delta t_{E}}\,S(\{n_{\omega}\})\left|\{n_{\omega}\}\right\rangle.
\end{equation}
Near the horizon the final state is given by (\ref{e:dm}). This is true for
both the in and out modes, so an observer stationed on either side of the
horizon would observe a thermal distribution of both ingoing and outgoing
particles.  As time passes, all of the ingoing particles will eventually
cross the horizon and be swallowed by the hole, whereas the outgoing particles
will propagate out to infinity where they can be detected at arbitarily late
times as a flux of thermal radiation at the Hawking temperature.

\newsection{Conclusion}

It was shown that the standard picture of black hole radiance is unchanged by
tunneling.  At late times, the hole radiates just as it would have had it been
formed from a classical collapse.  This makes sense if one thinks of Hawking
radiation as pair production.  The probability of tunneling is not affected
by the creation of a pair, since the pair has zero total energy. From this
point of view it is also clear that what happens at early times cannot
possibly affect the late time radiation, since the produced pairs only see the
late time geometry. The conventional derivation of radiance obscures this
point somewhat and it seems desirable to find an approach which makes
this feature manifest from the outset.
For the two systems considered here, and presumably this is true in general,
 the effect of the tunneling was to shift the distribution of any particles
 that were
 present before tunneling.  In the present case initial excitations were damped
because the final surface is rotated clockwise relative to the initial
surface.  A counterclockwise rotation would have led to amplification.
In \cite{Guth90} numerical investigations are quoted which show that the
rotation is always clockwise for the false vacuum bubble. One is led to
speculate whether this is a general phenomenon --- whether all tunneling
transitions lead to damping.

\newsection{Acknowledgements}

I would like to thank Frank Wilczek for many helpful discussions.

\newpage
\begin{center}
\bf{Figure Captions}\\[10mm]
\end{center}
1. A generic mirror potential.  The turning points for energy E
are indicated  \\[2mm]
2. The type (a) and (b) solutions.  The heavy lines represent the bubble
trajectory, and the dashed lines are the initial and final surfaces of the
tunneling solution.  In these figures, only the regions to the right of the
trajectory are of interest, as they are outside of the bubble.\\[2mm]
3. Bubble trajectory in Euclidean Schwarzschild space.

\end{document}